# SYMMETRIES AND THEIR BREAKING IN THE FUNDAMENTAL LAWS OF PHYSICS

## JOSE BERNABEU

Department of Theoretical Physics, University of Valencia,
and IFIC, Joint Centre Univ. Valencia-CSIC, E-46100 Burjassot

## Abstract
Symmetries in the Physical Laws of Nature lead to observable effects. Beyond the regularities and conserved magnitudes, the last decades in Particle Physics have seen the identification of symmetries, and their well defined breaking, as the guiding principle for the elementary constituents of matter and their interactions. Flavour SU(3) symmetry of hadrons led to the Quark Model and the antisymmetry requirement under exchange of identical fermions led to the colour degree of freedom. Colour became the generating charge for flavour-independent strong interactions of quarks and gluons in the exact Colour SU(3) local gauge symmetry. Parity Violation in weak interactions led to consider the Chiral Fields of fermions as the objects with definite transformation properties under the weak isospin SU(2) gauge group of the Unifying Electro-Weak SU(2)xU(1) Symmetry, which predicted novel weak neutral current interactions. CP-Violation led to three families of quarks opening the field of Flavour Physics. Time-Reversal-Violation has recently been observed with Entangled neutral mesons, compatible with CPT-invariance. The cancellation of gauge anomalies, that would invalidate the gauge symmetry of the quantum field theory, leads to Quark-Lepton Symmetry. Neutrinos were postulated in order to save the conservation laws of energy and angular momentum in nuclear beta decay. After the up's and down's on their mass, neutrino oscillations were discovered in 1998 opening a new era about their origin of mass, mixing, discrete symmetries and the possibility of global lepton-number violation through Majorana mass terms and Leptogenesis as the source of the matter-antimatter asymmetry in the Universe. The experimental discovery of quarks and leptons and the mediators of their interactions, with physical observables in spectacular agreement with this Standard Theory, is the triumph of Symmetries. The gauge symmetry is exact only when the particles are massless. One needs a subtle breaking of the symmetry, providing the Origin of Mass, without affecting the excellent description of the interactions. This is the Brout-Englert-Higgs Mechanism which produces the Higgs Boson as a remnant discovered at CERN in 2012. Open present problems are addressed with the search of New Physics Beyond-the-Standard-Model.

## Keywords
Flavour Families, Colour Charges, Gauge Symmetries, Chirality, Discrete Symmetries, Neutrinos, Spontaneous Breaking.

**Contents**

# 1. SYMMETRY AS GUIDING PRINCIPLE FOR PARTICLES AND INTERACTIONS

In ordinary life we observe symmetry of objects, like characteristic features of geometrical forms, material objects or biological bodies. The concept is related to the invariance of the object under definite transformations: One object is symmetric if, after a transformation is applied, the result remains the same, i.e., it remains "invariant". But we also observe Symmetry Breaking, which is particularly of interest when it is not a random effect but follows a definite pattern. In Fig. 1 we show the three-span arch of the FermiLab entrance, near Chicago, which appears perfectly symmetric when viewed from below, but has a calculated asymmetry from its other views. Symmetry and Symmetry Breaking are very important concepts in the field of elementary particle physics, however not referring to objects but to the Fundamental Laws of Physics.

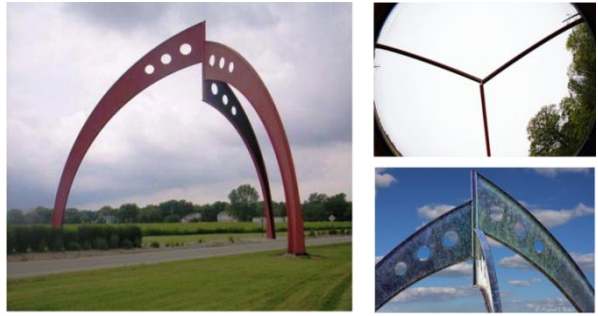

Fig. 1- Symmetry Breaking

We show here how Symmetry has acted as a guiding principle for both the existence of new particles and the formulation of interactions. One can claim that "Symmetry dictates Interaction", as stated by Yang. In Quantum Mechanics the Symmetry is implemented by a Unitary Transformation $\hat{U}$ acting on states and observables. If the Dynamics, described by the Hamiltonian $\hat{H}$, is Invariant under the transformation one has

$$[\hat{H}, \hat{U}] = 0 \qquad (1)$$

For continuous groups, under infinitesimal transformations generated by $\hat{G} = \hat{G}^\dagger$ one obtains immediately

$$\frac{d}{dt}\langle\hat{G}\rangle = i\langle[\hat{H}, \hat{G}]\rangle = 0 \qquad (2)$$

As $\hat{G}$ is Hermitian, it corresponds to an Observable that satisfies a Conservation Law if $\hat{U}$ is symmetry of $\hat{H}$. Well known examples are momentum for translations, angular momentum for rotations or charge for gauge symmetry. For local gauge symmetry the requirement of Invariance leads to a Covariant Derivative with a Mediator Field responsible of interactions. This is valid for either QED with the Abelian U(1) gauge group or non-Abelian gauge groups with the interaction field transforming as the adjoint representation.

In section 2 we develop the ideas leading from hadrons to quarks and the symmetries of strong interactions. In section 3 a parallel discussion is made for electroweak interactions starting from Parity Violation leading to the Standard Model with neutral currents and the need of charm plus the third family of bottom and top quarks, including Quark-Lepton Symmetry. Section 4 discusses Discrete Symmetries, with the observation and implications of the independent breaking of CP and T, as well as tests of CPT. In Section 5 the fascinating history of neutrino physics is summarized and the present open questions of CP-Violation in the lepton sector and Global-Lepton-Number Violation are discussed. Section 6 presents the Brout-Englert-Higgs Mechanism for the Origin of Mass, breaking the ElectroWeak Gauge Symmetry. Some Conclusions and Outlook in the field of Symmetries are given in Section 7.

## 2. QUARKS AND STRONG INTERACTIONS

The proliferation of non-strange and strange Hadrons in the 60's of the XX century led to the Eightfold Way of Gell Mann and Ne'eman with the use of the Flavour SU(3) symmetry. The fundamental representations $3, \bar{3}$ are the elementary building blocks for all higher-dimensional representations. In terms of their basis states, Mesons are $q - \bar{q}$ states $3 \times \bar{3}=1+8$, Baryons are q-q-q states $3 \times 3 \times 3 = 1 + 8_s + 8_a + 10$, with three quark q = u, d, s states. In Fig. 2 the octet and decuplet representations of Baryons are given in terms of third component of Isospin $I_3$ and hypercharge Y axes. According to the Gell Mann-Nishijima rule, the electric charge is $Q = I_3 + Y/2$, with Y = B + S, B the bayonic number and S strangeness.

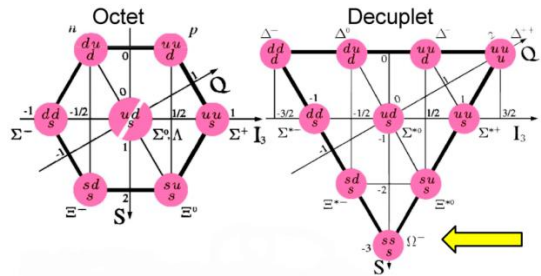

Fig. 2- Octet and Decuplet of Bayons

At the time of this formulation, the $\Omega^-$ had not been detected. Its later discovery was a great triumph of the whole scheme.

### 2.1. Are Quarks real?
For some time, however, the quark model for hadrons [1] was considered by the scientific community as a mere theoretical construct to describe the classification of hadrons in the SU(3) symmetry. The question was "Are Quarks real?". Since 1969, deep inelastic scattering experiments [2] at SLAC showed that the proton contained much smaller, point-like constituents and was therefore not an elementary particle. Physicists were reluctant to firmly identify these objects with quarks at the time, instead calling them "partons"—a term coined by Feynman. The partons that were observed at SLAC would later be identified as up and down quarks. Nevertheless, "parton" remains in use as a collective term for the constituents of hadrons (quarks, antiquarks and gluons). We do know at present that Leptons (electrons, muons, neutrinos) find partons in the proton with high momentum transfer events.

A "jet" is a narrow cone of hadrons produced by the hadronization of a parton. Jets were observed for the first time in the $e^+ e^-$ annihilation into hadrons at the SPEAR storage ring [3] and interpreted in terms of quarks. Quarks therefore exist, but they cannot propagate asymptotically. Quarks are then confined!

One of the reasons why the idea of real quarks was seen with scepticism was the problem of quarks with the exchange symmetry associated with the spin-statistics connection. It is easily realized with the $\Delta^{++}$ puzzle: The state $u^\uparrow u^\uparrow u^\uparrow$ with third component of total spin $S_3 = +3/2$ is evidently symmetric under exchange of flavour (u), quark spin ($s_3 = +1/2$) and space (L = 0) degrees of freedom of the three quarks!

If quarks are real fermions and satisfy the exchange symmetry, a new degree of freedom is necessary for quarks, the "colour" (r, g, b) being antisymmetric for its exchange in baryons. Precisely the singlet colour wave function

$$\Psi_c^{qqq} = \frac{1}{\sqrt{6}}(rgb - rbg + gbr - grb + brg - bgr) \tag{3}$$

is antisymmetric, so that qqq states exist, but these hadrons are colourless. We conclude that colour is confined, so that colourful quarks are confined. For the requirement of antisymmetry we need a number $N_c = 3$ of colours. Experimental evidence that $N_c = 3$ came from the interpretation of $e^+ e^- \rightarrow$ hadrons in terms of $q\,\bar{q}$ production, with a cross-section predicted to be proportional to $N_c$.

## 2.2. The Colour Charge
The Colour Charge appears as Generator of an exact SU(3)$_c$ Local Gauge Symmetry, leading to Colour Interaction of Quarks in the Fundamental Representation, mediated by 8 massless Gluons in the Adjoint Representation. This interaction is flavour-blind and only the quark mass terms break flavour independence. The origin of the quark mass terms should then be external to this QCD (Quantum ChromoDynamics) theory. The field tensor is covariant (A = 1, ..., 8) leading to self-interaction of the vector gluon field $A_\mu^A$ in the Lagrangian term $-\frac{1}{4} F_{\mu\nu}^A F^{A\mu\nu}$

$$F_{\mu\nu}^A = \partial_\mu A_\nu^A - \partial_\nu A_\mu^A - g_s f_{ABC} A_\mu^B A_\nu^C \tag{4}$$

with $g_s$ the quark-gluon coupling and $f_{ABC}$ the structure constants of the SU(3)$_c$ group. All coloured objects have strong interaction with gluons, so that quarks with gluons, gluons with themselves. Gluons have colour, so they are confined like quarks. Gluon Jets were first observed in the annihilation $e^+ e^- \rightarrow q\,\bar{q}\,g$ to three jets by the TASSO experiment [4] at the PETRA accelerator at the DESY laboratory.

The QCD coupling constant $\alpha_S = g_S^2/(4\pi)$ is dimensionless, therefore the classical field theory in the chiral (massless) limit is Scale Invariant. There is a Conformal Symmetry. However, in the perturbative treatment of the QCD quantum theory, predictions for observables are made in terms of the renormalized coupling $\alpha_s(\mu_R^2)$, which is a function of the renormalization scale. Taking it close to the momentum transfer $Q^2$, $\alpha_s$ ($Q^2$) indicates the effective strength of the interaction.

## 2.3. Asymptotic Freedom
The coupling runs with the renormalization scale $\mu_R^2$ and this running coupling satisfies the Renormalization Group Equation controlled by the QCD β($\alpha_s$) function. The 1 loop β function coefficient has contributions to the gluon self-energy from gluon self-couplings and fermion couplings with opposite signs. The dominance of the first term gives to QCD, distinct to QED, the property of **Asymptotic Freedom** [5]. The approximate analytic solution is

$$\alpha_s(\mu_R^2) = \left(b_0 ln\left(\mu_R^2/\Lambda^2\right)\right)^{-1}, \quad b_0 = (33 - 2n_f)/(12\,\pi) \tag{5}$$

with $n_f$ the number of flavour fermions and $\Lambda$ a constant of integration, representing the non-perturbative scale of QCD. The running coupling has been experimentally demonstrated with $\Lambda \sim 250$ MeV. The Dimensional Transmutation from $\alpha_s$ to $\Lambda$ is thus originated in the quantum Conformal Anomaly breaking the conformal symmetry. This $\Lambda$ is responsible of the nucleon mass and, as a consequence, the baryonic mass of the Universe!

## 3. CHIRALITY AND ELECTROWEAK INTERACTION

Parity Violation by Weak Interactions was postulated [6] in the 50's of the XX century to solve the puzzle of the different parities of the decay products of neutral Kaons. It was then observed in nuclear beta decay and later in charged pion decays.

Parity(P) $\vec{r} \rightarrow -\vec{r}$, Charge Conjugation(C) $q \rightarrow -q$ and Time Reversal(T) $\Delta t \rightarrow -\Delta t$ are Discrete Symmetries. In Fig. 3 we illustrate P and C transformations taking as Reference the observed $\pi^+ \rightarrow \mu^+ \nu_\mu$ decay

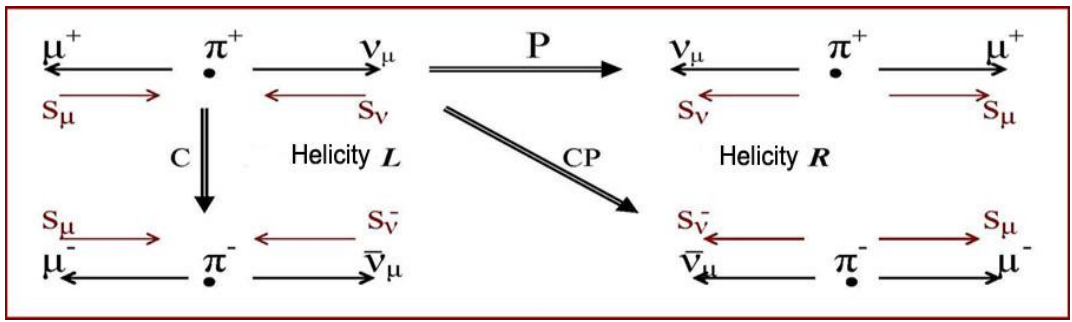

Fig. 3- The P, C and CP transformations in pion decays

Whereas the P-transformed and C-transformed processes do not exist for pion decays, the CP-transformed decay $\pi^- \rightarrow \mu^- \bar{\nu}_\mu$ is observed with the same decay rate. We conclude that Parity, as well as Charge Conjugation, is maximally violated, whereas CP is a good symmetry for pion decays.

We call a chiral phenomenon to one which is not identical to its mirror image. The spin component of a particle along its momentum may be used to define a handedness, or helicity. For massless fermions, the helicity is Lorentz invariant and this intrinsic property is the "Chirality". The invariance under Parity for a Dirac fermion state $\psi$ is called "Chiral Symmetry" and the transformation in Dirac space is implemented by the $\gamma_5$ Dirac matrix. Using Projectors, Left and Right chiral fermions, with definite chirality -1 and +1, are given by $\frac{1}{2}(1 - \gamma_5)\psi$, $\frac{1}{2}(1 + \gamma_5)\psi$. There are observables, like the vector and axial vector charges that conserve chirality of the fermions, whereas other observables, like the mass or dipole moments, connect the two chiralities.

In the Unified ElectroWeak Theory [7] based on the SU(2)$_L$ x U(1)$_Y$ gauge group, the fermion building blocks are not the Dirac fields $\psi$, but the Chiral Fields and the gauge group transformation distinguishes them: whereas the Left fields transform as Doublets under SU(2)$_L$, the Right Fields transform as Singlets under SU(2)$_L$. We say that this Unified Field Theory is a **Chiral Gauge Theory**.

The ElectroWeak gauge group SU(2)$_L$ x U(1)$_Y$ symmetry demands three gauge bosons W$_1$, W$_2$, W$_3$ of weak isospin from SU(2)$_L$ and the B boson of weak hypercharge Y from U(1)$_Y$. The gauge symmetry is here broken by the Mass terms and the Physical Fields with definite mass and charge are W±, γ, Z given by

$$\begin{pmatrix} \gamma \\ Z \end{pmatrix} = \begin{pmatrix} cos\theta_w & sin\theta_w \\ -sin\theta_w & cos\theta_w \end{pmatrix} \begin{pmatrix} B \\ W_3 \end{pmatrix}, \qquad M_z = \frac{M_W}{cos\theta_w} \qquad (6)$$

with θw the weak mixing angle. The theory predicts the existence of Weak Neutral Currents mediated by the Z boson and they were discovered [8] by the Gargamelle Bubble Chamber Collaboration at CERN in 1973 with muon neutrino interactions without muons in the final state. 10 years later, in 1983, the UA1 and UA2 experiments in the $S\bar{p}pS$ Collider at CERN discovered the massive W, Z Bosons as real particles reconstructed from their W⁺→ l⁺ νₗ, Z → l⁺ l⁻ [9] decays. These CERN discoveries established the triumph of the Standard Model of ElectroWeak Interactions.

### 3.1. GIM Mechanism: Need of Charm
With u, d, s quarks only, the Cabibbo d-s Mixing [10] in the Charged Weak Current leads, by the SU(2)L symmetry of the Standard Model, to Strangeness-Changing-Neutral Current at tree level implying, for example, fast KL → μ⁺ μ⁻ decay, against experiment. In 1970 Glashow-Iliopoulos-Maiani [11] solved this problem with an additional fourth quark flavour c completing two families of quark doublets

$$\begin{Bmatrix} u \\ d \end{Bmatrix} \begin{Bmatrix} c \\ s \end{Bmatrix}, \qquad U = \begin{pmatrix} cos\theta & -sin\theta \\ sin\theta & cos\theta \end{pmatrix} \qquad (7)$$

and interpreting the Cabibbo Mixing as a Unitary Mixing Matrix $U$ in d-s space, exhibiting the mismatch between weak eigenstates and mass eigenstates, with Charged Currents relative to both u and c quarks. SU(2)L then dictates that Neutral Currents are governed by U⁺ U = I, so they are Diagonal and Universal. Neutral Currents are Flavour-Conserving at tree level! At higher orders, Flavour-Changing-Neutral Currents can be induced from c-u mass difference. The KL →μ⁺ μ⁻ is suppressed -not only by higher orders- by the GIM additional factor (mc² - mu²)/Mw².

The discovery [12] of the $c\,\bar{c}$ J/ψ meson in 1974 at BNL and SLAC is coined as the November Revolution of particle physics. Charmed $c\,\bar{d}, c\,\bar{s}, cud$ ... hadrons were discovered later.

### 3.2. The third Family
The existence of a third family of quarks was predicted in 1973 by Kobayashi and Maskawa [13] in order to incorporate CP Violation in the Standard Model, to be discussed in Section 4, using a generalization of the Cabibbo Mixing for weak charged currents. The $b\,\bar{b}$ ϒ meson was discovered in 1977 at FermiLab and B mesons later.

The Top Quark is the most massive of all observed elementary particles. With a mass of 172.44 GeV/c², it weighs like an atom of tungsten! It decays by weak interaction t → b W with a lifetime of 5x10⁻²⁵ s. Such a short life is 1/20 of the timescale for Quark Hadronization, allowing "bare" quark studies with its entire spin density matrix in the production as well as in the decay.

The top quark was first indirectly "seen" with non-decoupling virtual quantum effects in $B^0 - \bar{B}^0$ Mixing [14] measured by UA1 and ARGUS in 1987, in the universal Z boson self energy [15] and in the specific Z $b\,\bar{b}$ vertex [16], the last two observed in the LEP experiments. The direct detection of top quarks was then made in 1995 at the $p\,\bar{p}$ Tevatron [17]. The p p

Collider LHC facility is at present a Top Quark Factory by means of its strong $g\,g \rightarrow t\,\bar{t}$ and weak $u\,\bar{d} \rightarrow t\,\bar{b}$ production mechanisms.

### 3.3. Gauge Anomalies: Quark-Lepton Symmetry

A Gauge Anomaly is a feature of quantum physics, a one-loop diagram, invalidating the gauge symmetry of a quantum field theory. All gauge anomalies must cancel out. Anomalies in gauge symmetries would destroy the required cancellation of unphysical degrees of freedom, such as a photon polarization in time direction.

Are gauge anomalies cancelled in the Standard Model? Anomalies appear in even D spacetime dimensions with **Chiral** fermions running in the loop with n = 1 + D/2 vertices. For D = 4, n = 3, it corresponds to Vector-Vector-Axial couplins! The condition for cancellation involves the particle content and the relations among their couplings [18]: the symmetrized trace over the generators of the gauge group vanishes

$$tr\big(\{\tau_i, \tau_j\}\tau_k\big) = 0 \qquad (8)$$

Such a cancellation operates within each family of Quarks and Leptons establishing an intriguing connection between the two sectors announcing a Grand Unification. For the three families required to incorporate CP Violation we then write the Symmetry between Quarks and Leptons.

$$\begin{Bmatrix} u \\ d \end{Bmatrix}\begin{Bmatrix} c \\ s \end{Bmatrix}\begin{Bmatrix} t \\ b \end{Bmatrix} \leftrightarrow \begin{Bmatrix} \nu_e \\ e \end{Bmatrix}\begin{Bmatrix} \nu_\mu \\ \mu \end{Bmatrix}\begin{Bmatrix} \nu_\tau \\ \tau \end{Bmatrix}$$

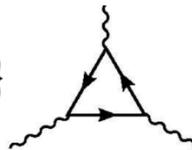

Fig. 4 - Quark-Lepton Symmetry requested for the cancellation of Eq. (8).

## 4. DISCRETE SYMMETRIES CP, T, CPT

As already discussed, the violation of Parity P and Charge Conjugation C had a profound impact in the way how the ElectroWeak Standard Model was formulated. In this section we present the state-of-the-art with the observed Violations of CP and T and the tests on CPT.

### 4.1. CP Violation

CP Symmetry would imply that the Laws of Physics should be invariant in form when a particle is interchanged with its antiparticle (C) while its spatial coordinates are inverted (P). For the neutral Kaon system with $\Delta S = 2$ Mixing $K^0 - \overline{K}^0$ by weak interactions, the physical states of definite mass and lifetime $K_L$, $K_S$ should be CP eigenstates leading to Conservation Laws: the decay $K_L \rightarrow \pi\,\pi$ should be forbidden. Its unexpected observation [19] in 1964 opened the entire new field of CP Violation in Flavour Physics.

Can CP Violation be described in the Standard Model? In 1973 Kobayashi and Maskawa discovered [13] such a possibility by breaking the CP symmetry in the Standard Model Lagrangian by means of the Cabibbo Mixing in the weak charged currents and enlarging the particle content of the theory. By going to, at least, three families of fermions the most general mismatch mixing matrix U between weak and mass eigenstates for d-s-b quarks contains a physical relative phase such that for antiquarks becomes its complex conjugate U*.

The precision measurements in the Kaon system with secondary Kaon beams [20] and at CPLEAR [21], and specially the observation of CPV in the $B_d$ system in the B-Factories [22, 23] in 2001 prompted the conviction that the Cabibbo-Kobayashi-Maskawa (CKM) mechanism of Quark Mixing with three families of fermions, is a good description of CPV in the SM. Recently, CPV in the charm system has been observed by the LHCb Collaboration [24] with the small magnitude as predicted by the SM. As seen, the origin of the Breaking is here advocated to be in the particle content of the theory.

All known laboratory experimental results on CP Violation for K, B and D physics are in agreement with the Unitary Mixing Matrix paradigm U (CKM) with three active families of quarks. In Fig. 5 the three families are written and the corresponding "Unitarity Triangle" relation for $B_d$-physics represented

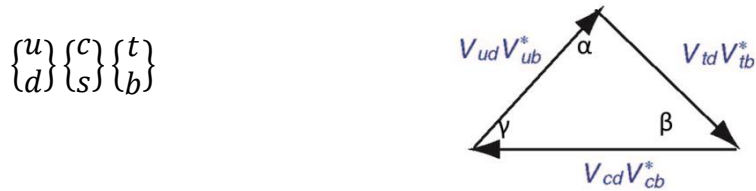

$$\begin{Bmatrix} u \\ d \end{Bmatrix} \begin{Bmatrix} c \\ s \end{Bmatrix} \begin{Bmatrix} t \\ b \end{Bmatrix}$$

Fig. 5- Three quark families and unitary triangle for $B_d$ physics

One should notice: i) the three upper u, c, t quarks have to be involved; ii) the three angles α, β, γ are CP violating observable phases, the first two involving the virtual $B^0 - \bar{B}^0$ mixing through the heavier t quark whereas γ is a signal of direct CP Violation in the decays to c and u quarks.

However, this Standard Model description of CP Violation is not enough to explain the Matter-Antimatter Asymmetry in the Universe!

### 4.2. Time Reversal Violation
A symmetry transformation T that changes the dynamics of a physical system into another with an inverted sense of time evolution is called Time Reversal (Reversal-in-Time). It is implemented in the space of states by an AntiUnitary Operator implying that its study has to be made with Asymmetries built under the exchange of in, out states.

The decay is an irreversible process indicating that a true TRV observable, needing a definite preparation and filtering of the appropriate initial and final particle state, looks impossible for transitions in the case of decaying particles. A bypass to this NO-GO argument was found [25] using entangled systems of unstable particles with the ingredients: i) The decay as a Filtering Measurement; ii) Entanglement implying the information transfer from the decayed particle to its living partner. For the entangled $B^0 - \bar{B}^0$ system produced by e⁺ e⁻ collisions at the Υ(4S) peak one may study the time dependence in meson transitions associated to the definite Flavour-CP eigenstate decay products. There are 2(Flavour) x 2(CP) x 2(time ordering) = 8 transitions of this kind which can be connected by different Separate Genuine T, CP, CPT Symmetry Transformations.

In Fig. 6 the experimental steps to measure the time dependent TRV Asymmetry for the $\bar{B}^0 \rightarrow B_-$ and $B_- \rightarrow \bar{B}^0$ meson transitions between Flavour and CP eigenstates are given where B₋ is the meson state filtered by the J/ψ Kₛ decay products.

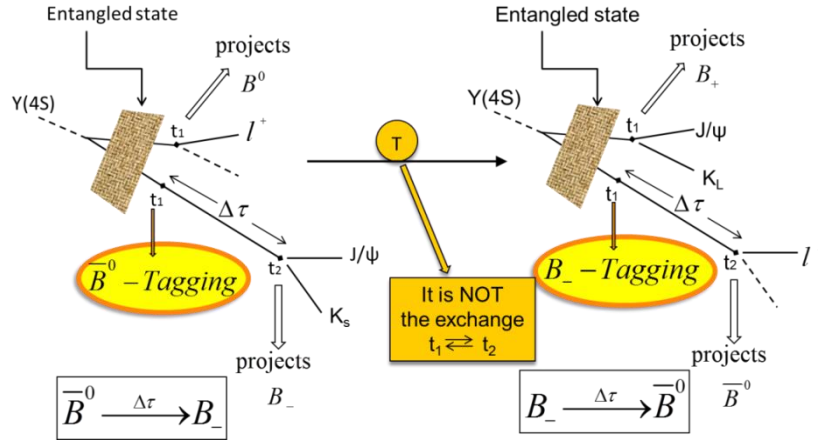

Fig. 6- Experimental steps to observe TRV in the entangled $B_d$-system

Using these concepts, the BABAR Collaboration observed [26] in 2012 a true TRV effect with 14 $\sigma$ significance. This discovery was a model-independent result, with the time-dependent form of the Intensity distribution as the only dynamical input, using the methodology previously defined in [27]. Its interpretation within the realm of the Weisskopf-Wigner approach for the effective Hamiltonian of the $B_d$-system, including CPT Violation, is full of interesting results [28] for the separation of genuine CPV and TRV observables.

For the neutral Kaon system, a search of TRV is going-on at the KLOE-2 experiment in the DAPHNE-Factory, using Entanglement as proposed in Ref. [29].

### 4.3 CPT

Although P, C, CP and T symmetries have been observed to be separately broken, there is no experimental evidence of CPT Violation. Its observation would imply a change of paradigm to beyond local QFT. There is, however, nothing against a possible breaking of CPT Invariance at the level of Quantum Mechanics. Most sensitive limits come from the neutral Kaon $K^0 - \overline{K}^0$ mass difference and the frequencies of the antihydrogen transitions.

I discuss Tests of the CPT symmetry using the semileptonic decays of neutral Kaons. Assuming the $\Delta S = \Delta Q$ rule, as implied by the description of flavoured mesons in terms of quarks, the allowed transitions are pictured by the diagrams

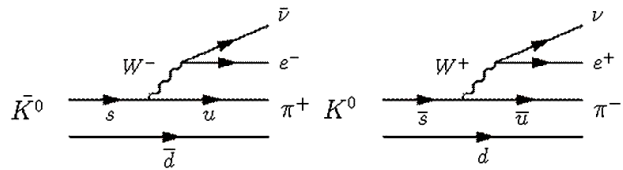

Fig. 7- Two decays are allowed according to elementary Quarks ($\Delta S = \Delta Q$ rule)

In the flavour states $K^0, \overline{K}^0$ the effective Hamiltonian in a Weisskopf-Wigner approach includes a non-diagonal CPV parameter and a diagonal CPTV parameter, so that the states with definite time evolution are

$$|K_s\rangle = \frac{1}{\sqrt{2\left(1+|\epsilon_s|^2\right)}}\left((1+\epsilon_s)|K^0\rangle + (1-\epsilon_s)|\bar{K}^0\rangle\right)$$

$$|K_L\rangle = \frac{1}{\sqrt{2(1+|\epsilon_L|^2)}}\left((1+\epsilon_L)|K^0\rangle - (1-\epsilon_L)|\bar{K}^0\rangle\right) \qquad (9)$$

where

$$\epsilon_{S/L} = \epsilon_K \pm \delta_K \qquad (10)$$

where "$\epsilon_K$" is the parameter describing CPV whereas "$\delta_K$" is the parameter describing CPTV. At CPLEAR, the preparation of an initial quantum state of flavour was used. With this tag, the following time-dependent combination of semileptonic decay rates was constructed as an asymmetry

$$\begin{cases} A_\delta(\tau) = \frac{\bar{R}+(\tau)-\alpha R_-(\tau)}{\bar{R}+(\tau)+\alpha R_-(\tau)} + \frac{\bar{R}-(\tau)-\alpha R_+(\tau)}{\bar{R}-(\tau)+\alpha R_+(\tau)} \\ R_{+(-)}(\tau) = R(K^0_{t=0} \to (e^{+(-)}\pi^{-(+)}v)_{t=\tau} \\ \bar{R}_{-(+)}(\tau) = R(\bar{K}^0_{t=0} \to (e^{-(+)}\pi^{+(-)}v)_{t=\tau} \\ \qquad \alpha = 1 + 4\Re\,\epsilon_L \end{cases} \qquad (11)$$

At times outside the interference region, i. e., longer than the short lifetime of $K_S$, this asymmetry becomes a pure CPT test and determined by the $\Re\,(\delta)$ parameter as

$$A_\delta(\tau \gg \tau_S) = 8\Re\delta \qquad (12)$$

It is worth noting that $\Re\,(\delta)$ does not need any absorptive part and it is determined by the difference of diagonal mass terms of $K^0$ and $\bar{K}^0$. The result of the measurement of the asymmetry (11) by the CPLEAR Collaboration [30] is

$$\Re\delta = (0.30 \pm 0.33 \pm 0.06) \times 10^{-3} \qquad (13)$$

where the first error is statistical and the second systematic.
Its interpretation in terms of the mass matrix elements of the $K^0 - \bar{K}^0$ system gives the upper bound

$$|m_{K^0} - m_{\bar{K}^0}|/m_K < 10^{-18} \qquad (14)$$

If we take as figure of merit this fractional difference between the masses of a particle and its antiparticle, the result (14) appears to be the best CPT Violation Limit.

At the $\phi$-Factory DAPHNE, the Entanglement between the two neutral Kaon states induced by their antisymmetric exchange allows a separate selection of observable asymmetries from either $K_L$ or $K_S$ states. The Charge Asymmetry from these states is given by

$$A_{S,L} = \frac{\Gamma(K_{S,L}\to\pi^-e^+v)-\Gamma(K_{S,L}\to\pi^+e^-\bar{v})}{\Gamma(K_{S,L}\to\pi^-e^+v)+\Gamma(K_{S,L}\to\pi^+e^-\bar{v})} = 2\,[\text{Re}\,(\epsilon_K) \pm \text{Re}\,(\delta_K) - \text{Re}\,(y)] \qquad (15)$$

where, in addition to the CPV and CPTV parameters in the Hamiltonian matrix, a possible CPTV in the flavour decay amplitudes of Figure 7, parameterised by "y", is included. As seen, the difference between these two Charge Asymmetries $A_S$ - $A_L$ is able to isolate CPTV effects in the diagonal matrix elements of the Mass Matrix.

The value of $A_L$ is well known from hadronic machines, with the best precise result from the KTeV Collaboration [31]

$$A_L = (3.332 \pm 0.058_{stat} \pm 0.047_{syst}) \cdot 10^{-3} \qquad (16)$$

The measurement of $A_S$ was undertaken in the first period of the KLOE experiment with the result [32]

$$A_S = (1.5 \pm 9.6_{stat} \pm 2.9_{syst}) \cdot 10^{-3} \qquad (17)$$

An important improvement of this result is expected at KLOE-2.

A direct test of CPT in neutral Kaon transitions has been proposed. EPR correlations at a $\Phi$-Factory can be exploited to study CPT-conjugated transition involving Flavour $K^0, \bar{K}^0$ states and the orthogonal $K_+, K_-$ states filtered by CP eigenstate decay products $\pi\pi$ and $\pi^0 \pi^0 \pi^0$.

For a CPT-symmetry test one has a priori 4 possible comparisons of transitions as given in Table 1

**CPT symmetry test**

| Reference | | $\mathcal{CPT}$-conjugate | |
|---|---|---|---|
| Transition | Decay products | Transition | Decay products |
| $K^0 \rightarrow K_+$ | $(\ell^-, \pi\pi)$ | $K_+ \rightarrow \bar{K}^0$ | $(3\pi^0, \ell^-)$ |
| $K^0 \rightarrow K_-$ | $(\ell^-, 3\pi^0)$ | $K_- \rightarrow \bar{K}^0$ | $(\pi\pi, \ell^-)$ |
| $\bar{K}^0 \rightarrow K_+$ | $(\ell^+, \pi\pi)$ | $K_+ \rightarrow K^0$ | $(3\pi^0, \ell^+)$ |
| $\bar{K}^0 \rightarrow K_-$ | $(\ell^+, 3\pi^0)$ | $K_- \rightarrow K^0$ | $(\pi\pi, \ell^+)$ |

Table 1.- Reference transitions and their CPT-transformed transitions.

For the most interesting $\Delta t$ region in the transition probabilities, with $\Delta t$ larger than the $K_S$ lifetime, one obtains [33]

$$R_2(\Delta t \gg \tau_S) = P\,[K^0 \rightarrow K_-]/P[K_- \rightarrow \bar{K}^0] = 1 - 4\,\mathrm{Re}(\delta)$$
$$R_4(\Delta t \gg \tau_S) = P\,[\bar{K}^0 \rightarrow K_-]/P[K_- \rightarrow K^0] = 1 + 4\,\mathrm{Re}(\delta) \qquad (18)$$

One cannot imagine better clearest tests of CPT-symmetry in transitions as those of Eq. (18). These tests are feasible at KLOE-2.

Up to now, we have considered CPTV-effects associated with the incompatibility of the CPT-operator with the Hamiltonian H, i.e., the non-invariance of H under CPT [CPT, H] ≠ 0. There is a possible second source of CPT-breaking induced by the quantum gravity vacuum: the ω-effect.

In presence of decoherence and CPT breaking induced by quantum gravity effects, the CPT operator is "ill-defined" and the definition of the particle-antiparticle states could be modified. This in turn could induce a weakening of the Entanglement [34] imposed by Bose statistics as an EPR correlation

$$|i\rangle \propto \left(|K^0\rangle|\bar{K}^0\rangle - |\bar{K}^0\rangle|K^0\rangle\right) + \omega\left(|K^0\rangle|\bar{K}^0\rangle + |\bar{K}^0\rangle|K^0\rangle\right)$$
$$\propto \left(|K_S\rangle|K_L\rangle - |K_L\rangle|K_S\rangle\right) + \omega\left(|K_S\rangle|K_S\rangle - |K_L\rangle|K_L\rangle\right) \qquad (19)$$

In some microscopic models of space-time foam [35] with stochastic fluctuations of defect recoils in the propagation of particles, the magnitude of the ω-parameter in Eq. (19) can reach values ω ∼ $10^{-4}$ - $10^{-5}$.

The maximum sensitivity to ω is expected for a symmetric decay pair of the entangled state f1= f2= $\pi^+ \pi^-$, which is CP-violating from the dominant antisymmetric state of Eq. (19). As a consequence, one gets a fantastic enhancement of the ω-effect from the "wrong" component in Eq. (19) as ω/ε, where ε is the small CPV parameter in K-physics. The higher luminosity of present DAPHNE and a good time resolution at KLOE-2 allow to reach an interesting sensitivity to this most fascinating ω-effect.

In the last years a new actor appeared: aggregate antimatter for tests of CPT. Antihydrogen experiments are pursuing very precise measurements of atomic transition frequencies. The CERN accelerator complex produces antiprotons through $p + p \rightarrow p + p + p + \bar{p}$ and, since 2000, the Antiproton Decelerator (AD) operates like all-in-one machine. The ELENA Project is expected to be in operation since 2021, with the Call of Proposals already done in 2019. Among the AD/ELENA experiments, the ALPHA Collaboration is involved in the 1S-2S spectroscopy in antihydrogen, and the ASACUSA Collaboration includes in its programme the antihydrogen hyperfine splitting (HFS).

In hydrogen the 1S-2S transition corresponds to a 2 photon channel with λ = 243 nm, measured with a precision Δ f/f = $10^{-14}$. The ground state hyperfine splitting corresponds to the famous λ = 21 cm line, with f = 1.4 GHz and measured with a precision Δ f/f = $10^{-12}$. The theoretical uncertainty, due to the proton structure, is however at the level of $10^{-6}$.

The first spectroscopy results with antihydrogen at AD were obtained by ALPHA for 1S-2S (∼ $10^{-12}$) [36] and for HFS (∼$10^{-4}$) [37]. New techniques in spectroscopy and antiproton beams promise higher accuracy for HFS by ASACUSA [38], and much more is expected to come with ELENA.

## 5. NEUTRINOS

The last two decades have seen a revolution in neutrino physics with the discovery of, and precision studies on, flavour oscillations in atmospheric [39], solar [40], reactor [41] and accelerator [42] neutrinos. These phenomena are interpreted in terms of non-vanishing masses and flavour mixing, the unitary PMNS matrix [43, 44] describing the mismatch between flavour and mass eigenstates. Global fits to all observable quantities provide better and better determination of the two mass differences $\Delta m_{21}^2$ and $|\Delta m_{31}^2|$, as well as the three mixing angles [45-47].

The most fundamental pending questions are the Dirac-Majorana confusion on the nature of neutrinos and their absolute mass scale, properties studied by means of non-accelerator methods, as well as the novel challenges for the next-generation neutrino flavour oscillation experiments like T2HK [48] and DUNE [49] using terrestrial accelerators. Above all, once known that the three mixing angles are non-vanishing [50–52], they should answer whether the lepton sector of elementary particles also incorporates CP violation, opening the door to concepts able to explain the matter-antimatter asymmetry of the Universe through leptogenesis [96] at higher energy scales. An additional open problem is the ordering of the neutrino mass spectrum, with the so-called either Normal or Inverted Hierarchies.

## 5.1. The Energy Crisis

Before 1930, the only subatomic particles known were the electron [53] and the proton [54]. Together with a still-developing quantum theory, these two elementary particles were enough to describe the atomic spectrum of hydrogen. The structure of atomic nuclei [55] was initially thought of in terms of bounded protons and electrons, especially motivated by the identification of $\beta$ radioactivity [56] with electron emission [57],

$$_Z^A X \rightarrow {}_{Z+1}^A Y + e^-$$ (20)

as is the case of $^{14}_6C$ decaying into $^{14}_7N$ with a half-life of about 5730 years. Conservation of energy and momentum in this two-body decay predicts a definite value for the emitted electron energy,

$$E_e = \frac{m_X^2 - m_Y^2 - m_e^2}{2m_X}$$ (21)

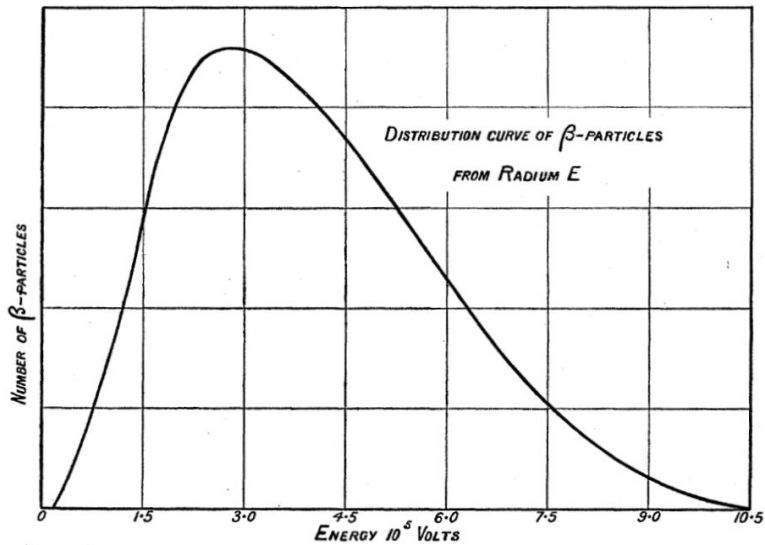

Fig. 8- β-decay electron energy spectrum [63].

Lise Meitner and Otto Hahn in 1911 [58] and Jean Danysz in 1912 [59] obtained the first hint that $\beta$ particles have a continuous spectrum, against the energy conservation rule leading to Eq. (21). Further evidence of the violation of energy conservation was given by James Chadwick in 1914 [60], measuring that the spectrum was continuous. In addition, molecular band spectra established [61] that the nuclear spin of Nitrogen-14 is 1, implying the violation of angular momentum conservation too in the process (20), which would thus involve the spins $0 \rightarrow 1 + \frac{1}{2}$. Dedicated studies in 1920–1927 by Charles Drummond Ellis and others [62 - 64] finally demonstrated beyond any doubt that the spectrum is continuous, as shown in Figure 8.

In a desperate reaction to these experimental facts, Niels Bohr postulated that conservation of energy was true in a statistical sense only [65]. These $\beta$ decay puzzles of non-conservation of energy and angular momentum were on the forefront of the open problems in the fundamental physics endeavour.

Indeed, the fact that the measured energy ranged from zero to a maximum given by its nominal value in Eq. (21), cannot be explained in this framework. Furthermore, the nuclear models with A protons and A-Z electrons were not able to explain the spin of the nuclei, which for e.g. $^{14}_7N$

should have been half-odd instead of the measured integer value. In this context, both conservation of energy and angular momentum seemed to be failing.

In 1930, Wolfgang Pauli wrote the famous letter of the 4th of December, addressed to the "Dear Radioactive Ladies and Gentlemen" participating in the Tübingen Conference: he proposed the existence of a hitherto unobserved spin-1/2 neutral particle with a small mass, no greater than 1% the mass of a proton, which he called a "neutron". In his words, this was a desperate remedy to save the exchange theorem of statistics and the law of conservation of energy. Assuming that this particle was emitted together with the electron in $\beta$ processes,

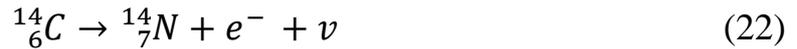

$$^{14}_{6}C \rightarrow {}^{14}_{7}N + e^- + v \tag{22}$$

the law of conservation of energy could successfully explain the spectrum in Figure 8, and angular momentum conservation could describe the $0 \rightarrow 1 + \frac{1}{2} + \frac{1}{2}$ transition too. Moreover, the presence of this hypothetical neutral particle in the nucleus in the same number as electrons could also solve the problem of nuclear spin.

Pauli excused his participation in the conference: «I cannot appear in Tübingen personally since I am indispensable here in Zurich because of a ball on the night of the $6^{th}$ to $7^{th}$ of December». The fact that this hypothetical particle had not been observed in $\beta$ decay forced it to be very weakly interacting, to the point that Pauli himself believed that it would never be observed. Unlike nowadays, when theoreticians have no problem in proposing particles whose observation is unfeasible, Pauli avoided claiming his solution in a scientific journal.

The nuclear spin problem was soon solved: James Chadwick discovered the neutron in 1932 [66]. Its spin 1/2 and mass similar to the proton mass opened the door to the understanding of the atomic nucleus in terms of protons and neutrons. Its heavy mass, however, made it clear that this particle could not be identified as Pauli's "neutron", and indeed the problems with conservation of energy and angular momentum in $\beta$ decays were still present.

The description of the subatomic world in terms of protons, electrons and neutrons lived a short life. A few months after Chadwick's discovery of the neutron, Carl D. Anderson took the photograph of a positron [67], finding the first evidence on the existence of antimatter. In 1934, Enrico Fermi incorporated Pauli's neutral particle, which he named "neutrino", together with all of these particles into his theory of $\beta^{\pm}$ decays [68], described as the subatomic processes

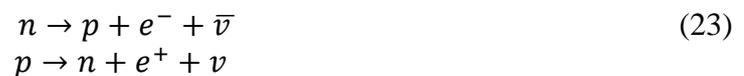

$$n \rightarrow p + e^- + \bar{v}$$
$$p \rightarrow n + e^+ + v \tag{23}$$

His theory successfully predicted the spectrum shapes, and provided a framework to calculate their lifetimes, supporting Pauli's proposed particle.

Using Fermi's theory, Hans Bethe and Rudolf Peierls estimated in 1934 [69] the mean absorption length of neutrinos in solid matter, a value larger than $10^{14}$ km: the expected cross section of neutrinos with matter was so low that they could traverse the Earth without any interaction. They concluded that «it is absolutely impossible to observe processes of this kind». Due to this low interaction rate, it was not until 1956 that the Cowan–Reines experiment [70] at the Savannah River Nuclear Reactor proved the existence of the neutrino. Antineutrinos created in the nuclear reactor by the $\beta$ decay were subsequently detected by the process

$$\bar{\nu} + p \rightarrow n + e^+ \tag{24}$$

in a large water tank. The signature of the process are the two 511 keV photons from the positron annihilation, in coincidence with the nuclear photon

$$n + {}^{108}Cd \rightarrow {}^{109}Cd + \gamma \tag{25}$$

due to the de-excitation of the ${}^{109}Cd$ nucleus after the neutron was captured by cadmium chloride molecules (CdCl$_2$) dissolved in the water.

Beyond the theoretical motivation for such a neutral particle proposed by Pauli and bolstered by Fermi, unquestionable evidence for its existence had been provided by Reines and Cowan. And so the neutrino was born.

## 5.2. Lepton Flavours

With the discovery of the muon [71, 72], a first problem appeared with its insertion in the Fermi theory of charged current weak interactions. Indeed, the muon was so unexpected that, regarding its discovery, Isidor Isaac Rabi famously quipped «Who ordered that?». A decade before the V-A theory [73], Bruno Pontecorvo discussed [74] the "universality" of Weak Interactions for nuclear $\beta$-decay processes together with those with muons and neutrinos. He introduced muon capture by nuclei and compared it with the probability for electron capture.

The idea of different neutrinos $\nu_\mu \neq \nu_e$ appeared published in 1959 by Pontecorvo [75] and, ever more important, in the proposal he made for the generation of a neutrino beam from pion decay [76].

In 1962, there was already a hint towards the different nature of the electron and muon neutrinos, coming from the $\mu \rightarrow e\gamma$ decay. An estimation of its branching ratio in the V-A theory with the W boson had yielded, if $\nu_\mu = \nu_e$, the value [77] $R_{th} \sim 10^{-4}$, whereas a limit $R_{exp} < 10^{-8}$ was found [78]. The Brookhaven experiment [79] was the first one involving high-energy neutrinos from pion decay. Using about $10^{14}$ muon antineutrinos from $\pi^-$, the experiment detected 29 events of the expected

$$\bar{\nu}_\mu + p \rightarrow \mu^+ + n \tag{26}$$

and no events of the forbidden

$$\bar{\nu}_\mu + p \rightarrow e^+ + n \tag{27}$$

The demonstration that $\nu_\mu \neq \nu_e$ was a great event in physics and thus two lepton families were completed:

1. $(\nu_e, e)$        2. $(\nu_\mu, \mu)$

After the discovery of the third charged lepton $\tau$ [80], the existence of (at least) a third neutrino was to be expected. The third neutrino $\nu_\tau$ was directly observed in 2000 in an experiment at Fermilab performed by the DONUT collaboration [81]. Using the Tevatron beam, a $\nu_\tau$ beam was mainly produced from $\tau$ decay, with the $\tau$ produced in the leptonic decay of a Ds meson. These $\nu_\tau$ were detected by means of the reaction

$$\nu_\tau + n \rightarrow p + \tau^- \tag{28}$$

by identifying the τ lepton as the only lepton created in the interaction vertex.

Two complementary methods for the determination of the number of neutrino species led to the highlighted LEP e+ e- Collider legacy at CERN [82]. The invisible width of the Z boson $\Gamma_{inv} = N_v\, \Gamma_v$ is not measurable directly, but the visible cross section depends on the number of neutrino species $N_v$ as

$$\sigma = \frac{4\pi s}{M_Z^2}\Gamma_{ee}\ \frac{\Gamma_Z - N_v\Gamma_v}{\left(s - M_Z^2\right)^2 + \Gamma_Z^2 M_Z^2} \qquad (29)$$

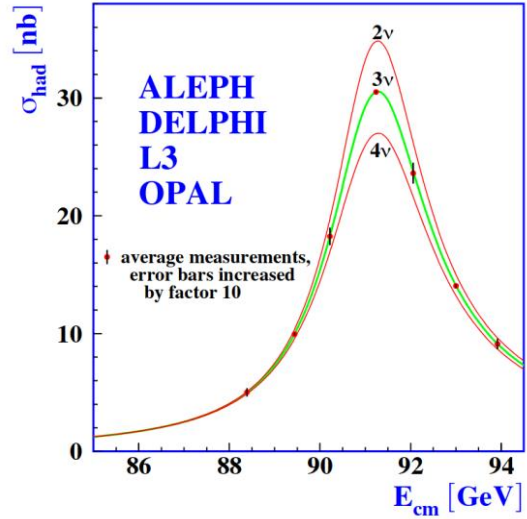

Fig. 9- Measurements of the hadron production cross-section around the Z resonance. The curves indicate the predicted cross-section for two, three and four neutrino species with SM couplings and negligible mass. [82]

The hadronic cross-section is shown in Figure 9, from which LEP could determine the existence of three (and only three) light active neutrinos.

An alternative at LEP was the measurement of the cross-section for $e^+e^- \to \gamma Z \to \gamma \nu \bar{\nu}$ detecting the photon plus nothing else. This cross section, normalized to γμμ, is a known function of $N_v$, and consistently reproduced the previous answers: there are three light active neutrinos.

### 5.3. Neutrino Mass and Mixing: Oscillations

Parity violation in processes involving neutrinos, like β decay, is automatic if neutrinos have a definite helicity. This fact led to the advent of the two-component theory -introduced by Hermann Weyl in 1929 [83] - if neutrinos are exactly massless, for which the Dirac equation

$$i\gamma^\mu\partial_\mu\, v_L(\text{x}) - m_v\, v_R(\text{x}) \ = \ 0 \qquad (30)$$

decouples. Non-conservation of parity was first observed in 1957 with Wu's experiment [84], measuring an asymmetry in the electron angular distribution emitted from the β decay of polarized $^{60}_{27}Co$ . In 1958, the celebrated Goldhaber experiment [85] proved that the neutrino helicity is -1, using conservation of angular momentum in a selected electron capture transition in $^{152}_{63}Eu$ leading to an excited nuclear state of $^{152}_{62}Sm$ —the measurement of the circular polarization of the de-exciting $\gamma$ ray fixed the neutrino helicity.

Even though this picture was consistent for massless neutrinos with definite helicity, the universal V-A theory of charged-current weak interactions was also formulated in 1958 [73], extending Fermi's theory of $\beta$ decay into a picture where left-handed chiral fields enter for all fermions, either neutrinos or not. Thus there is no rationale why neutrinos should be special and massless. Still, contrary to other fermions, neutrinos have no electric charge, so they present unique possibilities in the explanation of the origin of their masses. Whether these possibilities can be realized depends on the existence of a lepton charge that distinguishes $\nu$ $and$ $\bar{\nu}$. This is the most important open question even today!

Already in 1946 (!!), Pontecorvo asked the question whether antineutrinos produced from $\beta$ decay in reactors could produce electrons [86]. The negative result in 1959 of the Davis experiment [87] for

$$\bar{\nu} + {}^{37}_{17}Cl \rightarrow e^- + {}^{37}_{18}Ar \qquad (31)$$

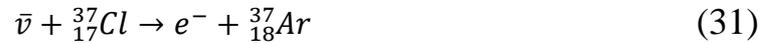

suggested the existence of definite lepton numbers with values $L_{e^-} = L_\nu = 1$ $and$ $L_{e^+} = L_{\bar{\nu}} =$ -1. Such a flavour charge would imply that neutrinos are not neutral particles, and so their mass could only be described with the same theories as all other fermions. The genius of Pontecorvo was the first to realize that this conclusion was referring to interacting neutrinos, and that there was still a possibility for neutrinos to acquire a Majorana mass.

The theory of Majorana particles had been introduced in 1937 by Ettore Majorana [88], who noticed that there are solutions of the Dirac equation for neutral particles with only two degrees of freedom: «the meaning of the Dirac equations is somewhat modified and there is no longer any reason to speak of negative-energy states nor to assume, for any other types of particles, especially neutral ones, the existence of antiparticles, corresponding to the "holes" of negative energy»

Forbidden for the other elementary fermions due to exact electric charge conservation, a Majorana mass for neutrinos is a priori allowed if the mass terms violate global lepton number by two units. In this case, the states of neutrinos with definite Majorana mass would be a linear superposition of weak interacting neutrinos with opposite lepton charge. Global lepton number would then be undefined for neutrinos with definite Majorana mass. Even more: one can have massive neutrinos with the active (left-handed) chiral component $\nu_L$ only and the sterile (right-handed) component $\nu_R$ is not needed. Contrary to Dirac fermions, Majorana fermions have only two degrees of freedom, the neutrino of left-handed chirality and its conjugate $\nu_L^c$. The states of definite mass and helicity, which are compatible observables, are the left-handed with a relative m/E component of the conjugate, and its orthogonal.

It is astonishing that the early ideas on neutrino mixing came through these reasonings. In 1957, Pontecorvo writes [43]: «If the theory of two-component neutrinos were not valid, and if the conservation law for "neutrino charge" took not place, neutrino-antineutrino transitions would be possible». In this statement one finds the two essential ingredients for oscillations: neutrino mass and mixing. He calculated [89] the survival probability of active neutrinos in a model of two Majorana neutrinos, one active, the other sterile, and suggested that the Cowan-Reines experiment should be repeated as function of the baseline for the detector. This kind of measurement in reactors was only performed for the first time in 2005 in the KAMLAND experiment [90].

In 1962, in the context of the Nagoya model of Baryons as bound states of neutrinos and "a new sort of matter" vector boson, Ziro Maki, Masami Nakagawa and Shoichi Sakata [44] introduced a mixing between two neutrinos to form the "true neutrinos" in these baryons. The objective was the explanation of the smallness of the leptonic decay rate of the hyperons, later explained by the Cabibbo mixing [10]. Although these ideas were not connected to neutrino oscillation phenomena, the concept of quantum mixings between neutrino states was there.

In 1967, after the discovery of the muon neutrino, Pontecorvo discussed [91] the phenomenology of neutrino oscillations in modern views, including the flavour transitions $v_e \to v_\mu$ and the Majorana transitions $v_e \to (v_e^c)_L$ and $v_\mu \to (v_\mu^c)_L$. Among other subjects, he applied this study to solar neutrino oscillations. At that time, Raymond Davis started his famous experiment on the detection of solar neutrinos [92]. They were detected via the observation of the reaction

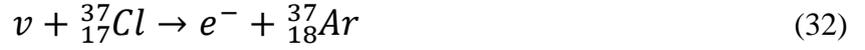

$$v + {}^{37}_{17}Cl \to e^- + {}^{37}_{18}Ar \qquad (32)$$

The results, measuring a $v_e$ flux significantly smaller than expected from solar models, created the solar neutrino problem.

The other grand historical problem in the neutrino field, the *atmospheric neutrino problem*, appeared in 1988 with the measurement of an anomalously small $v_\mu$ flux at Kamiokande [93]. The understood origin of atmospheric neutrinos from cosmic rays

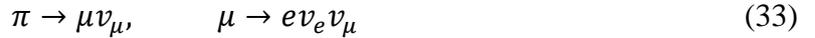

$$\pi \to \mu v_\mu, \qquad \mu \to e v_e v_\mu \qquad (33)$$

predicted a precise ratio of muon to electron neutrinos of 2:1, whereas the smaller muon neutrino flux observed lead to a ratio of 1.2.

These two problems were eventually solved with the consolidation of the oscillation paradigm. The discovery of neutrino oscillations in 1998 by Super Kamiokande [39], solving the atmospheric neutrino problem, together the later corroboration with solar neutrinos by SNO [40], solving the solar neutrino problem, marked the birth of the modern era in neutrino physics.

Historically, it is spectacular that the fundamental concepts of e - μ universality, flavour families, neutrino mixing and oscillations were understood in a period in which the prevailing idea was that of massless neutrinos. In the centenary of the birth of Pontecorvo, it was concluded [94] that it is fair to name the neutrino mismatch of flavour states with mass eigenstates as the PMNS Mixing Matrix.

## 5.4. Genuine CP Violation

Neutrino flavour oscillations have determined neutrino mass differences as oscillation frequencies and the mismatch mixings between flavour and mass eigenstates. Once known that the three mixing angles are non-vanishing, novel challenges include the search of CP Violation in the lepton sector. A recent constraint by T2K [95] on the allowed CP phase in the PMNS Mixing Matrix, at the level of 3σ, has been highly celebrated. This result is a prelude of the expectations for the next-generation experiments like T2HK [48] and DUNE [49]. The incorporation of CP Violation in the lepton sector would open the door to concepts able to explain the matter-antimatter asymmetry of the Universe through leptogenesis [96] at higher energy scales. A second open problem is the ordering of the neutrino mass spectrum, with the so-called Normal or Inverted Hierarchy.

Long baseline neutrino oscillation experiments propagate neutrinos from the source, created as muonic flavour, to the detector through the Earth mantle. The observation of CP violation needs an appearance experiment to a different flavour, and the "suppressed" transition to the electronic flavour is favoured. The corresponding CP violation asymmetry, defined in terms of the transition probabilities for neutrinos and antineutrinos is an odd function of L/E, with L the baseline and E the relativistic neutrino energy, if the propagation takes place in vacuum. Independent of particular theoretical frameworks, this observable is a bona-fide direct proof of CP violation. However, in actual experiments the propagation takes place in matter, which is CP-asymmetric, and a fake CP violation is originated through the different interaction of electron neutrinos and antineutrinos with the electron density of ordinary matter [97,98]. This complication in the quest for a direct evidence of fundamental CP violation is widely recognized, and some observables [99-101] have been tried for its separation. On the other hand, matter-induced terms are welcome in order to obtain information on the neutrino mass hierarchy. The restoration of the idea that a direct evidence of symmetry violation means the measurement in a single experiment of an observable odd under the symmetry has been solved recently [102,103]. The Concept exploited is based on the fact that genuine and matter-induced CP violation have opposite behaviors [104] under the other discrete symmetries of Time-Reversal T and CPT: whereas genuine CP violation is odd under T and even under CPT, the matter effect is T-even and CPT-odd. Although they are well separated in the effective Hamiltonian, in general they are not in the experimental observables and, in particular, in the CP asymmetry. The ideal way to solve this problem would be the independent measurement of T-odd [105] and CPT-odd asymmetries, but this route requires sources of electron neutrinos and antineutrinos above the muon mass energies, which is at present unavailable for accelerator facilities. As an alternative, the solution consists in disentangling these two components, genuine and matter-induced CP violation, in the observable CP asymmetry.

A Disentanglement Theorem [102] has been proved for any flavour transition. Neutrino oscillation CP violating asymmetries in matter have two disentangled components: i) a CPT-odd T-invariant term, non-vanishing if there are interactions with matter; ii) a T-odd CPT-invariant term, non-vanishing if there is genuine CP violation. As function of the baseline, these two terms are distinct L-even and L-odd observables to separately test (i) matter effects sensitive to the neutrino hierarchy and (ii) genuine CP violation in the neutrino sector. For the golden $\nu_\mu \rightarrow \nu_e$ channel, the different energy distributions of the two components provide a signature of their separation. At long baselines, they show oscillations in the low and medium energy regions, with zeros at different positions and peculiar behaviour around the zeros. A Magic Energy E= (0.91±0.01) GeV exists at L= 1300 km with vanishing CPT-odd component and maximal genuine CP asymmetry proportional to sinδ, with δ the weak CP phase. For energies above 1.5 GeV, the sign of the CP asymmetry discriminates the neutrino hierarchy.

In the experimental region of existing and planned terrestrial accelerator neutrinos, it is proved [103] that, at medium baselines -T2K and T2HK-, the CPT-odd component is small and nearly δ-independent, so it can be subtracted from the experimental CP asymmetry as a theoretical background, provided the hierarchy is known. At long baselines -DUNE-, on the other hand, one finds that (i) a Hierarchy-odd term in the CPT-odd component dominates the CP asymmetry for energies above the first oscillation node, and (ii) the fake CPT-odd term vanishes, independent of the CP phase δ and the hierarchy, at the magic energy E= 0.91 GeV (L/1300 km) near the second oscillation maximum, where the genuine T-odd term is almost maximal and proportional to sinδ. A measurement of the CP asymmetry in these energy regions would thus provide separate information on (i) the neutrino mass ordering, and (ii) direct evidence of genuine CP violation in the lepton sector.

## 5.5. Global Lepton Number

Knowing that neutrinos are massive, the most fundamental problem is the determination of the nature of neutrinos with definite mass: are they either four-component Dirac particles with a conserved global lepton number L, distinguishing neutrinos from antineutrinos, or two-component truly neutral (no electric charge and no global lepton number) self-conjugate Majorana particles [88]? For Dirac neutrinos, like quarks and charged leptons, their masses can be generated in the Standard Model of particle physics by spontaneous breaking of the gauge symmetry with the doublet Higgs scalar, if there were additional right-handed sterile neutrinos. But the Yukawa couplings would then be unnaturally small compared to those of all other fermions. One would also have to explain the origin of the global lepton number avoiding a Majorana mass for these sterile neutrinos. A Majorana $\Delta L = 2$ mass term, with the active left-handed neutrinos only, leads to definite-mass neutrinos with no definite lepton charge. However, there is no way in the Standard Model to generate this Majorana mass, so the important conclusion in fundamental physics arises: Majorana neutrinos would be an irrefutable proof of physics beyond the Standard Model. Due to the Majorana condition of neutrinos with definite mass being their own antiparticles, Majorana neutrinos have two additional CP-violating phases [106-108] beyond the Dirac case.

Neutrino flavour oscillation experiments cannot determine the fundamental nature of massive neutrinos. In order to probe whether neutrinos are Dirac or Majorana particles, the known way has been the search of processes violating the global lepton number L. The dificulty encountered in these studies is well illustrated by the so-called confusion theorem [109, 110], stating that in the limit of zero mass there is no difference between Dirac and Majorana neutrinos. As all known neutrino sources produce highly relativistic neutrinos (except for the present cosmic neutrino background in the universe), the $\Delta L = 2$ observables are highly suppressed. Up to now, there is a consensus that the highest sensitivity to small Majorana neutrino masses can be reached in experiments on the search of the L-violating neutrinoless double-β decay process ($0\nu\beta\beta$). Dozens of experiments around the world are seeking out a positive signal, and the most sensitive limits are obtained by GERDA-II [111] in $^{76}$Ge, CUORE [112] in $^{130}$Te and KAMLAND-Zen [113] in $^{136}$Xe. An alternative to $0\nu\beta\beta$ is provided by the mechanism of neutrinoless double electron capture ($0\nu$ECEC) [114], which actually corresponds to a virtual mixing between a nominally stable parent Z atom and a daughter (Z-2)* atom with two electron holes. The experimental process is the subsequent X-ray emission and it becomes resonantly enhanced when the two mixed atomic states are nearly degenerate. The search of appropriate candidates [115] satisfying the resonant enhancement is being pursued with the precision in the measurement of atomic masses provided by traps. The process can be stimulated [116] in XLaser facilities. The $2\nu$ECEC decay, allowed in the Standard Model, has recently been observed for the first time by the XENON Collaboration [117] in $^{124}$Xe. This last process, contrary to the case of $2\nu\beta\beta$ for searches of $0\nu\beta\beta$, is not an irreducible background for $0\nu$ECEC when the resonance condition is satisfied.

Recently a novel idea [118] has been developed following a different path to the search of $\Delta L = 2$ processes. It is based on having a pair of virtual non-relativistic neutrinos of definite mass, whose quantum distinguishability is different for Dirac and Majorana nature due to the lepton charge. Such a physical situation is apparent in the long-range force mediated by two neutrinos at distances near its range of microns, well beyond the atomic scale. Neutral aggregate matter has a weak charge and there is a 3 x 3 Hermitian matrix of six coherent charges for its interaction with two neutrinos of equal or different masses. The effective two-neutrino potential near its range leads to observable absolute neutrino mass and the Dirac/Majorana distinction via two effects: a different r-dependence and a violation of the weak equivalence principle.

The search for the elusive neutrino properties of the absolute neutrino mass and the Global Lepton Number symmetry is of primary importance and, although experimentally challenging, worth to be pursued using all alternative ways.

## 6. THE BROUT-ENGLERT-HIGGS MECHANISM

The Standard Model of particle physics contains as elementary constituents the three families of fermions with the Quark-Lepton Symmetry. Their interactions appear as a requirement of the local Gauge Symmetries $SU(3)_c$ x $SU(2)_L$ x $U(1)_Y$ generated by the three charges of colour, weak isospin and weak hypercharge. The last two combine to the electric charge for $U(1)_{em}$. These interactions operate as Exchange Forces with the Mediators Gluon with m = 0, but confined, Photon with m = 0 and the massive $W^{\pm}$, Z bosons. The Standard Model not only predicted new particles and interactions, but its agreement with all precision experimental results of detailed Observables in the last decades is impressive. However, these gauge symmetries are exact only in the massless limit, against the facts in nature. One should have a very subtle mechanism for the Origin of Mass without affecting the interactions, responsible of the $SU(2)_L$ x $U(1)_Y$ gauge symmetry breaking into $U(1)_{em}$. This is the Brout-Englert-Higgs Mechanism [119].

The Spontaneous ElectroWeak Symmetry Breaking (SEWSB) is based on the possibility that a symmetric Law of Physics can lead to asymmetric solutions. One should be aware that a Quantum Field Theory needs for its precise definition not only the Lagrangian (the physical law) but also the Quantum Vacuum, the lowest energy state from which particles are created and annihilated. SEWSB means that the physical law is symmetric and the vacuum is asymmetric. How?

The spacetime is filled with a "medium", a complex scalar field with the interaction being like a "mexican hat"

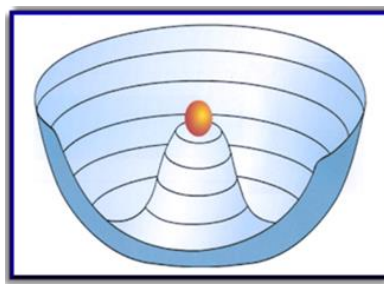

Fig. 10- Interaction of the complex scalar field.

This behaviour is obtained from a negative "mass square" quadratic term plus a positive quartic term. We observe that, instead of a unique symmetric lowest energy state, there are many possible vacua and one choice breaks the symmetry. This "spontaneous symmetry breaking" could be called a hidden symmetry because the results are independent of the chosen vacuum.

The physical particle created from the new vacuum is the Higgs Boson, a remnant of the Brout-Englert-Higgs Mechanism, hence its importance. There is a cristal clear signature of the Higgs particle: its coupling to all particles, including to itself, is proportional to their mass, a property that breaks the gauge symmetry. The Origin of Mass comes from the asymmetry of the new vacuum.

On 4 July 2012, the ATLAS and CMS experiments at CERN's Large Hadron Collider announced [120] they had each observed a new particle in the mass region around 125 GeV. In Fig.11 we show these original data together with the comparison of the measured partial decay rates to different channels to the expected theoretical predictions in the Standard Model.

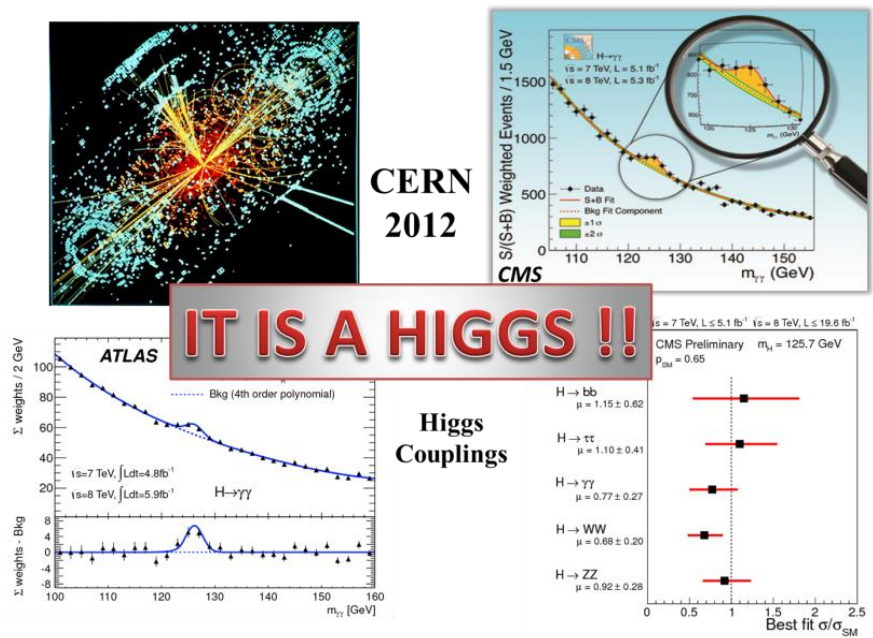

Fig. 11- Production and Decay of the Higgs boson observed in the ATLAS and CMS experiments.

As seen, the couplings are consistent with hose expected for a Higgs particle.

On 8 October 2013 the Nobel prize in physics was awarded jointly to François Englert and Peter Higgs "for the theoretical discovery of a mechanism that contributes to our understanding of the origin of mass of subatomic particles, and which recently was confirmed through the discovery of the predicted fundamental particle, by the ATLAS and CMS experiments at CERN's Large Hadron Collider."

### 6.1 The Higgs Boson gauge and Yukawa couplings

In 2020 many Higgs properties have already been established, among them its mass 125 GeV, spin/parity $0^+$, width < 1 Gev (direct) and < 0.015 Gev (indirect) and the observed direct couplings to Vector Bosons, $\tau$ leptons and top quarks. This is well illustrated in Fig. 12 using the combined ATLAS + CMS experimental results

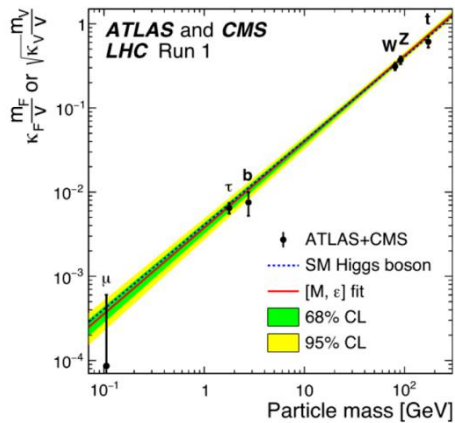

Fig. 12- The observed direct couplings of the Higgs to vector bosons W, Z and fermions.

As seen, all measurements are compatible with SM predictions. The decay channel $H \to b\,\bar{b}$, the unique final state to measure the coupling with down-type quarks, is most relevant. It has the largest branching fraction: 58 % for $M_H$ = 125 GeV. The problem, however, is the huge QCD background. Hence **the solution** of studying the production of $V\,H(b\,\bar{b})$ by $q\,\bar{q}$ with V = W, Z. The case Z H has an additional g-g produced amplitude. The leptonic decay channels of $V$ are preferred W → l ν and Z → l l, νν. ATLAS and CMS have more than 5σ separate observations of the H → b b Decay confirming the SM result on the Yukawa coupling to b's.

The SM Lagrangian, as summarized in Fig. 13, contains gauge propagation and self-interaction in the first line, fermion propagation and interaction with gauge fields in the second, the last two lines involving the Higgs Boson. The Yukawa couplings -in red- and the gauge couplings -in green- are present in Higgs Production and in Higgs Decays as advertised by the arrows.

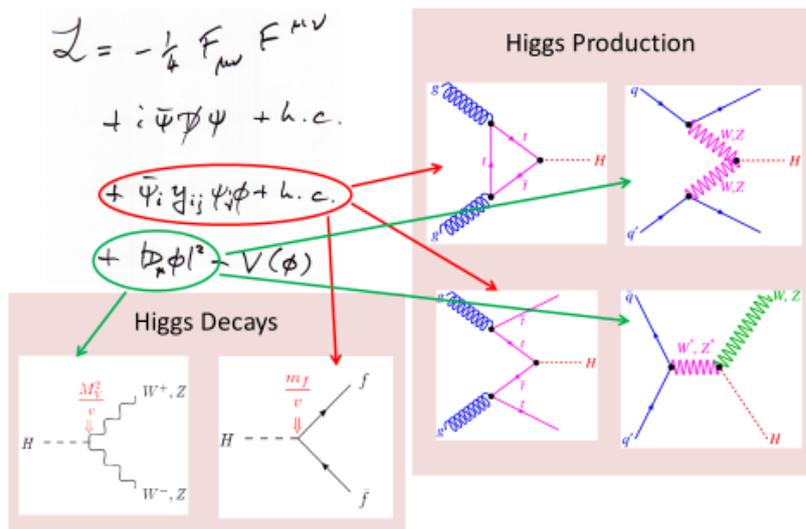

Fig. 13 - The Higgs Lagrangian and its test by means of Higgs Production and Higgs Decay.

We may conclude that there is strong evidence that the gauge and Yukawa couplings of the H-particle at 125 GeV correspond to a Higgs. However, the SM Higgs potential V($\phi$) has not been experimentally explored up to now! We need new accessible observables.

## 6.2 The Higgs Potential

The SM scalar sector of the theory is constructed for a doublet under SU(2) of Complex Fields

$$\phi = \begin{pmatrix} \phi^+ \\ \phi^0 \end{pmatrix}$$

Besides gauge and Yukawa couplings, the **HIGGS POTENTIAL** is given in Eq. (34)

$$V(\phi) = \mu^2 \phi^\dagger \phi + \lambda (\phi^\dagger \phi)^2 \tag{34}$$

and it is invariant under local gauge transformations.

For $\mu^2 < 0$, $\lambda > 0$, the potential energy density V($\phi$) has a minimum at $\phi^\dagger \phi = \frac{-\mu^2}{2\lambda} = \frac{v^2}{2}$

This is a degenerate vacuum, seen in the Mexican hat of Fig. 10, as a consequence of gauge invariance. Let us choose the direction $\phi_0 = \frac{1}{\sqrt{2}} \begin{pmatrix} 0 \\ v \end{pmatrix}$, $\phi(x) = \frac{1}{\sqrt{2}} \begin{pmatrix} 0 \\ v + H(x) \end{pmatrix}$, so that H(x) is the remnant neutral scalar field defined from the new vacuum: the Higgs.

The choice of the direction implies three broken global symmetries leading, therefore, to three massless Goldstone Bosons: two charged and one neutral. Physically they are associated to the longitudinal degrees of freedom of the massive W$^\pm$ and Z bosons. When transforming the fields of definite mass the local gauge symmetry appears broken. This is the "Spontaneous Symmetry Breaking"(SSB) mechanism. The symmetry is still there, hidden in the invariance of the Lagrangian under the transformation of the original non-physical fields, and it is reminded with the independence of physics with the particular choice of the direction responsible of the SSB.

After the SSB, the SM Higgs potential -in terms of the physical Higgs field- becomes

$$V(H) = \frac{m_H^2}{2} H^2 + \lambda_3^{SM} v H^3 + \lambda_4^{SM} H^4 \tag{35}$$

where $m_H^2 = -2\mu^2$, $\lambda_3^{SM} = \frac{m_H^2}{2v^2}$, $\lambda_4^{SM} = \frac{m_H^2}{8v^2}$

The Higgs vacuum expectation value (VEV) v = 246 GeV is related to the Fermi coupling constant G$_F$ measured in muon decay. Thus we have a self-coupling $\lambda_3^{SM} = 0.13$

Eq. (35) for the Higgs potential was never probed and it is waiting its measurement in order to complete the triumph of the SM for explaining the secrets of nature. **The self-couplings of the Higgs boson are the keystone**.

## 7. CONCLUSIONS AND OUTLOOK

The three sectors of the Standard Model -Strong, ElectroWeak and Higgs- represent a tribute to the concept that Symmetry, and Symmetry Breaking, is the Guiding Principle for Particles and Interactions and the Origin of Mass. The particle content has been induced by Discrete Symmetry Breaking for fermions, opening the field of Flavour Physics for Quarks and Leptons.

We have emphasized the role of different definite patterns for the breaking of symmetries, like
- Mass terms are incompatible with both gauge and chirality symmetries.
- Quantum loop Anomalies break conformal symmetry for vector theories and gauge symmetry for chiral field theories.
- The Particle Content of the theory controls the breaking of Discrete Symmetries CP and T.
- A gauge Asymmetric Vacuum leads to spontaneous symmetry breaking with hidden gauge symmetry and explaining the Origin of Mass for elementary particles.

What next? There are theoretical and observational reasons for searching Beyond-Standard-Model Physics at LHC experiments and in other facilities. I list some of them:
- Why the quantization of electric charge, Dual QED, Magnetic Monopoles
- The principle of "Threeality" in fundamental physics.
- The Hierarchy Problem for scalars, Supersymmetry.
- Grand Unification, p-decay.
- Neutrino Mass, Mixing, CPV, Global Lepton Number.
- Charged Lepton Flavour Violation.
- Baryon Asymmetry of the Universe, Leptogenesis.
- Dark Matter
- Dark Energy

Most ideas tackling these points are linked to the paradigm that symmetries will continue to be the guiding principle for fundamental physics. Among the Discrete Symmetries, CPT is protected by the "CPT-Theorem" in quantum field theory formulated in Minkowski spacetime with interactions satisfying Lorentz invariance, locality and unitarity. But there is nothing at the level of quantum mechanics which forbids to have CPT-violation and there are sound quantum gravity arguments in favour of this ultimate symmetry breaking.

**ACKNOWLEDGEMENT**

I would like to thank the organizers of this Special Issue for the invitation to present these ideas on the role of symmetries in fundamental physics. This research has been supported by the FEDER/MCIyU-AEI Grant FPA2017-84543-P and Generalitat Valenciana Project GVPROMETEO 2017-033